\documentclass[12pt]{article}

\usepackage{amsmath}
\usepackage{amssymb}
\usepackage{amsthm}
\usepackage{mathrsfs}
\usepackage[T1]{fontenc}
\usepackage{enumerate}
\usepackage{color}

\def\cond{condition}

\def\tfn{transformation}

\def\dd{Drinfel'd double}

\def\4diml{four-dimensional}

\def\-1{^{-1}}
\def\half{\frac{1}{2}}

\def\wwt{\widetilde}

\def\pltd{Poisson--Lie T-dualit}

\def\cf{{\mathcal {F}}}

\newcommand{\rmd}{{\mathrm{d}}}
\newcommand{\nn}{\nonumber}
\newcommand{\Lie}{\pounds}
\newcommand{\gLie}{\hat{\pounds}}
\newcommand{\unit}{\mathbf{1}}

\title{Classification of six-dimensional Leibniz algebras ${\mathcal E}_3$}
\author{Ladislav Hlavat\'y\footnote{hlavaty@fjfi.cvut.cz}
\\ {\em Faculty of Nuclear Sciences and Physical Engineering,}
\\ {\em Czech Technical University in Prague,}
\\ {\em Czech Republic}
}
\begin{document}
\maketitle

\begin{abstract}
Leibniz algebras ${\mathcal E}_n$ were introduced as algebraic
structure underlying U-duality. Algebras ${\mathcal E}_3$ derived
from Bianchi three-dimensional Lie algebras are classified here. Two
types of algebras are obtained: Six-dimensional Lie algebras that
can be considered extension of semi-Abelian four-dimensional \dd\
and unique extensions of non-Abelian Bianchi algebras. For all of
the algebras explicit forms of generalized frame fields are given.
\end{abstract}



\section{Introduction}
Extensions of  \pltd ies to non-perturbative symmetries of string
theories are so called U-dualities (for review see
e.g.\cite{oberspiolin98}). Algebraic structures underlying
U-duality were suggested in \cite{sakatani:uduality} and
\cite{malthom} as Leibniz algebras ${\mathcal E}_n$ obtained as
extensions of $n$-dimensional Lie algebra defining non-symmetric
product $\circ$ in $[n+n(n-1)/2]$-dimensional vector space\footnote{for $n\leq 4$} that
satisfies Leibniz identity
\begin{equation}
\label{LIdentity} X\circ(Y\circ Z))=(X\circ Y)\circ Z+Y\circ(X\circ Z).
\end{equation}
In those papers examples of these Leibniz algebras derived from
two-dimensional  and four-dimensional Lie algebras are
given. Goal of the present note is to write down all algebras that
can be derived from three dimensional Lie algebras whose
classification given by Bianchi is well known.

Namely, let $(T_a,T^{a_1a_2}),\ a,a_1,a_2\in {1,\ldots,n},\
 T^{a_2a_1}=-T^{a_1a_2}$ is a basis of $[n+n(n-1)/2]$-dimensional vector space.
 The algebra product given in \cite{sakatani:uduality} is
\begin{align}
\begin{split}
 T_a\circ T_b &= f_{ab}{}^c\,T_c \,,
\\
 T_a\circ T^{b_1b_2} &= f_a{}^{b_1b_2c}\,T_c + 2\,f_{ac}{}^{[b_1}\, T^{b_2]c} \,,
\\
 T^{a_1a_2}\circ T_b &= -f_b{}^{a_1a_2 c}\,T_c + 3\,f_{[c_1c_2}{}^{[a_1}\,\delta^{a_2]}_{b]} \,T^{c_1c_2}\,,
\\
 T^{a_1a_2}\circ T^{b_1b_2} &= -2\, f_d{}^{a_1a_2[b_1}\, T^{b_2]d}\,,
\end{split}
\label{eq:En-algebra}
\end{align}
where $f_{ab}{}^c$ are structure coefficients of $n$-dimensional Lie algebra and $f_a{}^{b_1b_2b_3}=f_a{}^{[b_1b_2b_3]}$.
Moreover, bilinear forms on ${\mathcal E}_n$ are defined
\begin{align}
 \langle T_a,\, T^{b_1b_2}\rangle_c = 2!\,\delta^{[b_1}_{a}\,\delta^{b_2]}_c \,,\qquad
 \langle T^{a_1a_2},\, T^{b_1b_2}\rangle_{c_1\cdots c_4} = 4!\,\delta^{a_1}_{[c_1}\,\delta^{a_2}_{c_2}\,\delta^{b_1}_{c_3}\,\delta^{b_2}_{c_4]}\,.
\label{eq:section}
\end{align}

\section{Bianchi-Leibniz algebras}
We are going to classify Leibniz algebras
${\mathcal E}_3$ derived from three-dimensional Lie algebras. In
this case $f_a{}^{b_1b_2b_3}=f_a{}\,\varepsilon^{b_1b_2b_3}$ where
$\varepsilon$ is totally antisymmetric Levi-Civita symbol.
Non-vanishing bilinear forms are
$$ \langle T_1,\, T^{12}\rangle_2 =\langle T_1,\, T^{13}\rangle_3 =\langle T_2,\, T^{23}\rangle_3 =1, $$
$$ \langle T_2,\, T^{12}\rangle_1 =\langle T_3,\, T^{13}\rangle_1 =\langle T_3,\, T^{23}\rangle_2 =-1. $$

First of all we shall show that for dimension three the Leibniz
identities are satisfied only for unimodular Lie
Algebras\footnote{This is not true in general as can be shown
explicitely for dimension four or by two-dimensional example in
\cite{sakatani:uduality}.},  i.e. $f_{ab}{}^b=0$. Indeed, Leibniz
identity
$$ T^{23}\circ(T_1\circ T_1)=(T^{23}\circ T_1)\circ T_1+T_1\circ(T^{23}\circ T_1)$$
and definitions \eqref{eq:En-algebra} give
$$ 0=2\,(f_{12}{}^2+f_{13}{}^3)^2\,T^{23}=2\,(f_{1b}{}^b)^2\,T^{23}$$
and similarly for cyclic permutation of $(1,2,3).$

Next point in our computations is the well known classification of
3--dimensional real Lie algebras.
Non--isomorphic Lie algebras can be divided into eleven classes,
traditionally known as Bianchi algebras. Their Lie algebra products
are (see e.g. \cite{Landau})
\begin{equation}
[X_1,X_2]=-a X_2+n_3 X_3, \; [X_2,X_3]=n_1 X_1, \; [X_3,X_1]=n_2 X_2 + a X_3,
\label{bian}
\end{equation}
where the parameters $a,n_1,n_2,n_3$ have the values given in the Table \ref{table1}.
\begin{table}
\begin{center}
\begin{tabular}{|c|r|r|r|r|}
\hline
{\bf Class} & $a$ & $n_1$ & $n_2$ & $n_3$ \\
\hline
B$\mathbf 1$& 0 & 0 & 0 & 0 \\
B$\mathbf 2$ & 0 & 1 & 0 & 0 \\
B$\mathbf 3$ & 1 & 0 & 1 & -1 \\
B$\mathbf 4$ & 1 & 0 & 0 & 1 \\
B$\mathbf 5$ & 1 & 0 & 0 & 0 \\
B${\mathbf 6_0}$ & 0 & 1 & -1 & 0 \\
B${\mathbf 6_a} \, (a>0,a \neq 1)$ & $a$ & 0 & 1 & -1 \\
B${\mathbf 7_0}$ & 0 & 1 & 1 & 0 \\
B${\mathbf 7_a}  \, (a>0)$ & $a$ & 0 & 1 & 1 \\
B$\mathbf 8$ & 0 & 1 & 1 & -1 \\
B$\mathbf 9$ & 0 & 1 & 1 & 1 \\
\hline
\end{tabular}\caption{Bianchi algebras }
\label{table1}
\end{center}
\end{table}
Unimodular Bianchi algebras are those with $a=0$, i.e. B$\mathbf 1$ (Abelian), B$\mathbf 2$ (Heisenberg), B${\mathbf 6_0}$ (Euclidean), B${\mathbf 7_0}$ (Poincare),
B$\mathbf 8$ (so(2,1)), and B${\mathbf 9}$ (so(3)).

Inserting \eqref{bian} and \eqref{eq:En-algebra} into Leibniz identities  \eqref{LIdentity} we get
\begin{equation}\label{njfk}
 n_jf_k=0,\quad j,k=1,2,3 .
\end{equation}
This can be shown inspecting e.g. identities \eqref{LIdentity} for
$$X=T_1,\ Y=T^{23},\ Z=T^{12}, $$ and $$X=T_1,\ Y=T^{23},\ Z=T^{13}. $$
We get
$$ n_2f_1\,T^{13}+n_2f_2\,T^{23}=0,$$
$$ n_3f_1\,T^{12}+n_3f_3\,T^{23}=0,$$
so that
\begin{equation}\label{n2f1}
n_2f_1=0,\quad n_2f_2=0,\quad n_3f_1=0,\quad n_3f_3=0.
\end{equation}
By cyclic permutation  of $(1,2,3)$ we get \eqref{njfk} and it is easy to check
 that these \cond s are sufficient for satisfaction of all Leibniz identities
 \eqref{LIdentity}.
Solution of \cond s \eqref{njfk} is either $n_j=0,\ j=1,2,3$ or $f_k=0,\ k=1,2,3$.  It means that
we get two types of Bianchi-Leibniz algebras.

The first type are algebras
depending only on $f_k$ with products
\begin{align}
\begin{split}
 T_a\circ T_b &= 0 \,,
\\
 T_a\circ T^{b_1b_2} &= f_a\,\varepsilon^{b_1b_2c}\,T_c  \,,
\\
 T^{a_1a_2}\circ T_b &= -f_b\,\varepsilon^{a_1a_2 c}\,T_c \,,
\\
 T^{a_1a_2}\circ T^{b_1b_2} &= -2\, f_d\,\varepsilon^{a_1a_2[b_1}\, T^{b_2]d}\,.
\end{split}
\label{abelianEn}
\end{align}
It is rather easy to check that this product is antisymmetric so
that they are six-dimensional Lie algebras. The simplest one is Abelian where all $n_j=0$ and $f_k=0$. If at least one of $f_k$ is not zero then by linear \tfn\ from $E_3=SL(3)\times SL(2)$ we can
achieve $f_1=1,\,f_2=f_3=0$ so that 
\begin{align}
\begin{split}
 [T_a,T_b] &= 0 \,,
\\
 [T_1\circ T^{b_1b_2}] &= 2\,\varepsilon^{b_1b_2c}\,T_c  \,,
\\
 [T^{23}, T^{12}]= 2\, T^{12},\quad &[T^{23}, T^{13}]= 2\, T^{13} \,.
\end{split}
\label{abelianEn1}
\end{align}

The Bianchi-Leibniz algebras of the second type depend only on $n_j$
whose values are given in the Table \ref{table1}. It means that they
are in one to one correspondence with the unimodular Bianchi
algebras. Their products are
\begin{align}
\begin{split}
 T_a\circ T_b &= [T_a,T_b] \,,
\\
 T_a\circ T^{b_1b_2} &=\delta_a^{b_1}\varepsilon_{ab_2c}\,n_{b_2}\,T^{b_1c} -\delta_a^{b_2}\varepsilon_{ab_1c}\,n_{b_1}\,T^{b_,c}  \,,
\\
 T^{a_1a_2}\circ T_b &=0\,,
\\
 T^{a_1a_2}\circ T^{b_1b_2} &=0\,.
\end{split}
\label{BianchiEn}
\end{align}
Explicit forms of products $ T_a\circ T^{b_1b_2}$ are
$$ T_1\circ T^{12}=-T_3\circ T^{23}=n_2\,T^{13},$$
$$ T_1\circ T^{13}=T_2\circ T^{23}=-n_3\,T^{12},$$
$$ T_2\circ T^{12}=T_3\circ T^{13}=n_1\,T^{23}.$$

Maximal isotropic algebras in both types of algebras are generated by $\{T_1,T_2,T_3\}$, $\{T^{12},T^{13},T^{23}\}$ and $\{T_1,T^{23}\}$, $\{T_2,T^{13}\}$, $\{T_3,T^{12}\}$.

As mentioned in \cite{sakatani:uduality}, under some \cond s we can
choose a subalgebra of dimension $2(n-1)$ of the Leibnitz algebra
${\mathcal E}_n$ that is Lie algebra of \dd. Leibniz algebra then
can be considered as an extension of \dd\ of dimension $2(n-1)$.
Namely, if we can decompose the generators $\{T_a\}$ as
$\{T_{\dot{a}},T_z\}$ and $\{T^{ab}\}$ as
$\{T^{\dot{a}\dot{b}},\,T^{\dot{a} z}\}$ ($\dot{a}=1,\dotsc,n-1$) so
that
\begin{align}
 f_{ab}{}^{z} = 0\,,\quad f_{az}{}^{b} = 0 \,,\quad f_z{}^{b_1b_2b_3} = 0 \,,\quad f_{\dot{a}}{}^{\dot{b}_1\dot{b}_2\dot{b}_3}=0\,,
\label{solvable}
\end{align}
then the subalgebra spanned by
\begin{align}
 (T_{\dot{A}})\equiv (T_{\dot{a}},\,T^{\dot{a}}) \qquad (T^{\dot{a}}\equiv T^{\dot{a}z})
\label{eq:Drinfeld-embedding1}
\end{align}
becomes Lie algebra of \dd\ with the bilinear form
\begin{align}
 \langle T_{\dot{a}},\, T^{\dot{b}}\rangle := \langle T_{\dot{a}},\, T^{\dot{b}}\rangle_z = \delta^{\dot{b}}_{\dot{a}}\,.
\end{align}

However for $n=3$, the conditions \eqref{solvable} and unimodularity
are satisfied only for Abelian Bianchi algebra B${\mathbf 1}$. It
means that Leibniz algebras \eqref{abelianEn} can be considered
extensions of Lie algebras of four-dimensional  \dd s
\cite{hlasno:pltdm2dt} generated by $T_1,T_2,T^{13},T^{23}$ and
$$ [T_1,T_2]=0,\quad [T^{13},T^{23}]=-f_1\,T^{13}-f_2\,T^{23},$$
$$ \langle T_1,\, T^{13}\rangle =\langle T_2,\, T^{23}\rangle=1. $$

\subsection{Generalized frame fields}
Now we will present explicit forms of the so called generalized frame
fields $E_A{}^I$ required to satisfy
\begin{align}\label{Lieder EAEB}
 \gLie_{E_A} E_B{}^I = - {\cf}_{AB}{}^C\, E_C{}^I,
\end{align}
where ${\cf}_{AB}{}^C$ are structure constants of the
Bianchi-Leibnitz algebras, $$
A,B,C,I\in(1,2,3,\{1,2\},\{1,3\},\{2,3\}). $$ The generalized Lie
derivative can be expressed by 
\begin{align}
 \bigl(\gLie_V W^I\bigr) =
 \begin{pmatrix}
 \Lie_v w^i \\
 \frac{(\Lie_v w_2 - \iota_w \rmd v_2)_{i_1i_2}}{\sqrt{2!}} \end{pmatrix},
\end{align}
generalized vectors $V$ and $W$ are parameterized as
\begin{align}
 V^I =
 \begin{pmatrix}
 v^i \\
 \frac{v_{i_1i_2}}{\sqrt{2!}}
 \end{pmatrix},\qquad
 \end{align}
$v_2=\half v_{ij}\rmd x^{i}\wedge\rmd x^j$ and similarly for $W$ and $\wwt w$.

Generalized frame fields $E_A{}^{I}$ have block triangular form
\begin{align}\label{EAI}
 E_A{}^I &= \begin{pmatrix} {e_a}^i & 0 \\E^{a_1a_2i} &{E^{a_1a_2}}_{i1,i_2} \end{pmatrix}= \begin{pmatrix} {e_a}^i & 0 \\ -\tfrac{\Pi^{a_1a_2 b}\,e_b^i}{\sqrt{2!}} & {r_{[i_1}}^{[a_1}\,{r_{i_2]}}^{a_2]} \end{pmatrix} ,
\end{align}
where ${e_a}^{i}$ are components of right-invariant vector fields  with respect to Bianchi groups, ${r_i}^{a}$ are components of right-invariant 1-forms and $\Pi^{a_1a_2b}$ is the so called Nambu-Poisson tensor that
in the dimension three is $\pi(x_1, x_2, x_3)\, \varepsilon^{a_1a_2b}$ where $\varepsilon$ is the Levi-Civita symbol.

From the formula \eqref{EAI} follows explicit form of matrices
$E_A{}^{I}$ for the first type algebras \eqref{abelianEn}, namely
\begin{align}
 E_A{}^I &= \begin{pmatrix} \unit_3 & 0 \\ \Pi_3 &\half\, \unit_3 \end{pmatrix} \end{align}
where $\unit_3$ is three-dimensional unit matrix,
\begin{align}
\Pi_3 =\begin{pmatrix} 0 & 0 &-\pi(x_1, x_2, x_3) \\0 &\pi(x_1, x_2, x_3)& 0 \\-\pi(x_1, x_2, x_3) & 0 & 0   \end{pmatrix},
\end{align}
and $\pi(x_1, x_2, x_3)=f_1x_1+f_2 x_2+f_3 x_3 +const$ where $f_a$
are constants appearing in \eqref{abelianEn}.

Explicit form of matrices $E_A{}^{I}$ for the second type algebras
\eqref{BianchiEn} is
\begin{align}\nn
 E_A{}^I &= \begin{pmatrix} M_1 & 0 \\ 0 &\half\, M_2 \end{pmatrix} \end{align}
where
\begin{align}\nn
 M_1 &=\left(
\begin{array}{ccc}
 1 & 0 & 0 \\
 0 & 1 & 0 \\
 -n_1 x_2 & x_1 & 1 \\
\end{array}
\right),
\end{align}
\begin{align}\nn
 M_2 &= \left(
\begin{array}{ccc}
 1 & -x_1 & - n_1 x_2 \\
 0 & 1 & 0 \\
 0 & 0 & 1 \\
\end{array}
\right)\end{align}
for Bianchi algebras B$\mathbf 2$, B${\mathbf 6_0}$, B${\mathbf 7_0}$.
\begin{align}\nn
 M_1 &= \left(
\begin{array}{ccc}
 1 & 0 & 0 \\
 -\sinh x_1\tanh x_2& \cosh x_1& \sinh x_1\,\text{sech}\ x_2\\
 -\cosh x_1\tanh x_2& \sinh x_1& \cosh x_1\,\text{sech}\ x_2\end{array}
\right),
\end{align}
\begin{align}\nn
 M_2 &=  \left(
\begin{array}{ccc}
 \cosh x_1& - \sinh x_1\cosh x_2& - \sinh x_2\cosh x_1\\
- \sinh x_1&  \cosh x_1\cosh x_2&  \sinh x_1\sinh x_2\\
 0 & 0 &  \cosh x_2\\
\end{array}
\right)\end{align}
for Bianchi B$\mathbf 8$.
\begin{align}\nn
 M_1 &= \left(
\begin{array}{ccc}
 1 & 0 & 0 \\
 \sin x_1\tan x_2& \cos x_1& -\sin x_1\sec x_2 \\
 -\cos x_1\tan x_2& \sin x_1& \cos x_1\sec x_2\\
\end{array}
\right),
\end{align}
\begin{align}\nn
 M_2 &=  \left(
\begin{array}{ccc}
  \cos x_1& - \sin x_1\cos x_2& - \sin x_2\cos x_1\\
  \sin x_1&  \cos x_1\cos x_2& - \sin x_1\sin x_2\\
 0 & 0 &  \cos x_2\\
\end{array}
\right)\end{align}
for Bianchi B$\mathbf 9$.

\section{Malek-Thompson modification of frame fields}
In the paper \cite{malthom} another algebra underlying U-duality was
presented starting from a more general form of the frame field
\begin{align}
 E_A{}^I &= \begin{pmatrix} {e_a}^i & 0 \\ -\tfrac{\Pi^{a_1a_2 b}\,e_b^i}{\sqrt{2!}} & \alpha\, {r_{[i_1}}^{[a_1}\,{r_{i_2]}}^{a_2]} \end{pmatrix}
\end{align}
that for $\alpha=1$ coincides with \eqref{EAI}. This ansatz for
$\alpha=e^{Z_ax^{a}}$ leads to modification of the algebra \eqref{eq:En-algebra} to the form \cite{sakatani:uduality}
\begin{align}
\begin{split}
 T_a\circ T_b &= f_{ab}{}^c\,T_c \,,
\\
 T_a\circ T^{b_1b_2} &= f_a{}^{b_1b_2c}\,T_c + 2\,f_{ac}{}^{[b_1}\, T^{b_2]c} - Z_a\,T^{b_1b_2} \,,
\\
 T^{a_1a_2}\circ T_b &= -f_b{}^{a_1a_2 c}\,T_c + 3\,f_{[c_1c_2}{}^{[a_1}\,\delta^{a_2]}_{b]} \,T^{c_1c_2} + 3\,Z_{[b}^{\vphantom{a_1}}\,\delta^{a_1a_2}_{c_1c_2]}\,T^{c_1c_2}\,,
\\
 T^{a_1a_2}\circ T^{b_1b_2} &= -2\, f_d{}^{a_1a_2[b_1}\, T^{b_2]d}
\end{split}
\label{eq:Enalgebra2}
\end{align}
This modification enables to define the algebras $\mathcal E_3$ also for the non-unimodular Bianchi algebras.

Leibniz identities for this generalized algebra in case $n=3$
require $Z_a=-{f_{ab}}^{b}$ so that they admit also the
non-unimodular Bianchi algebras B${\mathbf 3}$, B${\mathbf 4}$,
B${\mathbf 5}$, B${\mathbf 6_a}$, B${\mathbf 7_a}$ beside those
given in the preceding Section. In these cases $n_1= 0$ and
${f_a}^{bcd}=0,\ Z_1=-{f_{1b}}^{b}=2a,\ Z_2=Z_3=0$. In other words,
non-vanishing products of these algebras  are
\begin{align}
\begin{split}
[T_1,T_2]=-a T_2+n_3 T_3,& \; [T_2,T_3]=0, \; [T_3,T_1]=n_2 T_2 + a T_3\,,
\\
 T_1\circ T^{12} &=-a\, T^{12}+\,n_2\,T^{13}  \,,
\\
 T_1\circ T^{13} &=-a\, T^{13}-\,n_3\,T^{12}  \,,
\\
 T_2\circ T^{23} &=-a\, T^{13}-\,n_3\,T^{12}  \,,
\\
 T_3\circ T^{23} &=a\, T^{12}-\,n_2\,T^{13},
\end{split}
\label{BianchiEn2}
\end{align}
where the values of parameters $a,n_2,n_3$ are given in the Table 1.
\subsection{Twisted generalized frame fields}
It is understandable that for  structure constants  ${{\cf}'}_{AB}{}^C$ of the modified algebra \eqref{BianchiEn2}
the generalized frame fields $E_A{}^I$ of the form \eqref{EAI} do not satisfy relations \eqref{Lieder EAEB}.
For example, the formula \eqref{EAI} for Bianchi algebra
B${\mathbf 5}$ gives
$${E_A}^{I} = diag(1,e^{x_1},e^{x_1},\half ,e^{-x_1},\half
,e^{-x_1},\half ,e^{-2x_1})  $$ and  $$\gLie_{E_1} E_6{}^I
=(0,0,0,0,0,e^{-2x_1}).$$ But, as follows from \eqref{BianchiEn2},
$$ {{\cf}'}_{1,6}{}^C {E_C}^{I}=(0,0,0,0,0,0).$$

To satisfy relations
\begin{align}\label{Lieder EAEB twist}
 \gLie_{{{E}'}_A} {{{E}'}_B}^{I}= - {{\cf}'}_{AB}{}^C\, {{{E}'}_C}^{I},
\end{align}
where ${{\cf}'}_{AB}{}^C$ are structure coeficients of the algebra \eqref{BianchiEn2} the generalized frame fields must be modified by a twist matrix $T$ as ${{{E}'}_A}^{I} ={E_A}^{J}{T_J}^{I}$ \cite {sakatani:uduality,malthom}.
It concerns the non-unimodular Bianchi algebras
B${\mathbf 3}$, B${\mathbf 4}$, B${\mathbf 5}$, B${\mathbf 6_a}$, B${\mathbf 7_a}$ and the twist matrix is
\begin{align}\label{twistmtx}
T_J{}^I &= \begin{pmatrix} \unit_3 & 0 \\0& e^{2a\,x_1}\unit_3
\end{pmatrix}. \end{align} where $a$ is given  in the Table 1.

Corresponding twisted generalized frame fields  are
$$ {{{E}'}_A}^{I} =e^{ x_1}\left(
\begin{array}{cccccc}
 e^{ -x_1}& 0 & 0 & 0 & 0 & 0 \\
 0 & \cosh x_1& \sinh x_1& 0 & 0 & 0 \\
 0 & \sinh x_1& \cosh x_1& 0 & 0 & 0 \\
 0 & 0 & 0 & \frac{1}{2} \cosh x_1& -\frac{1}{2} \sinh x_1& 0 \\
 0 & 0 & 0 & -\frac{1}{2} \sinh x_1& \frac{1}{2} \cosh x_1& 0 \\
 0 & 0 & 0 & 0 & 0 & \frac{1}{2} e^{- x_1} \\
\end{array}
\right)$$
for B${\mathbf 3}$,

$$ {{{E}'}_A}^{I} =\left(
\begin{array}{cccccc}
 1 & 0 & 0 & 0 & 0 & 0 \\
 0 & e^{x_1} & -e^{x_1} x_1 & 0 & 0 & 0 \\
 0 & 0 & e^{x_1} & 0 & 0 & 0 \\
 0 & 0 & 0 & \frac{e^{x_1}}{2} & 0 & 0 \\
 0 & 0 & 0 & \frac{1}{2} e^{x_1} x_1 & \frac{e^{x_1}}{2} & 0 \\
 0 & 0 & 0 & 0 & 0 & \frac{1}{2} \\
\end{array}
\right)$$
for B${\mathbf 4}$,

$$ {{{E}'}_A}^{I} =\left(
\begin{array}{cccccc}
 1 & 0 & 0 & 0 & 0 & 0 \\
 0 & e^{x_1} & 0 & 0 & 0 & 0 \\
 0 & 0 & e^{x_1} & 0 & 0 & 0 \\
 0 & 0 & 0 & \frac{e^{x_1}}{2} & 0 & 0 \\
 0 & 0 & 0 & 0 & \frac{e^{x_1}}{2} & 0 \\
 0 & 0 & 0 & 0 & 0 & \frac{1}{2} \\
\end{array}
\right)$$
for B${\mathbf 5}$,

$$ {{{E}'}_A}^{I} =e^{a\, x_1}\left(
\begin{array}{cccccc}
 e^{-a\, x_1}& 0 & 0 & 0 & 0 & 0 \\
 0 & \cosh x_1& \sinh x_1& 0 & 0 & 0 \\
 0 & \sinh x_1& \cosh x_1& 0 & 0 & 0 \\
 0 & 0 & 0 & \frac{1}{2} \cosh x_1& -\frac{1}{2} \sinh x_1& 0 \\
 0 & 0 & 0 & -\frac{1}{2} \sinh x_1& \frac{1}{2} \cosh x_1& 0 \\
 0 & 0 & 0 & 0 & 0 & \frac{1}{2}e^{-a\, x_1}\\
\end{array}
\right)$$
for B${\mathbf 6_a}$,

$$ {{{E}'}_A}^{I} =e^{a\, x_1}\left(
\begin{array}{cccccc}
 e^{-a\, x_1} & 0 & 0 & 0 & 0 & 0 \\
 0 & \cos x_1& -\sin x_1& 0 & 0 & 0 \\
 0 & \sin x_1& \cos x_1& 0 & 0 & 0 \\
 0 & 0 & 0 & \frac{1}{2} \cos x_1& -\frac{1}{2} \sin x_1& 0 \\
 0 & 0 & 0 & \frac{1}{2} \sin x_1& \frac{1}{2} \cos x_1& 0 \\
 0 & 0 & 0 & 0 & 0 & \frac{1}{2} e^{-a\, x_1} \\
\end{array}
\right)$$
for B${\mathbf 7_a}$.

These twisted generalized frame fields then satisfy relations
\eqref{Lieder EAEB twist}.

\section{Conclusions} We have classified six-dimensional Leibniz
algebras \eqref{eq:En-algebra} and \eqref{eq:Enalgebra2} starting
from Bianchi classification of three-dimensional Lie algebras.

Up to linear transformations from $E_3=SL(3)\times SL(2)$ we have obtained seven inequivalent
algebras \eqref{eq:En-algebra}. Two of them, obtained from Abelian Lie
algebra B${\mathbf 1}$, are six-dimensional Lie algebras
\eqref{abelianEn} that can be considered extensions of semi-Abelian
four-dimensional \dd. The other five are unique Leibniz extensions
\eqref{BianchiEn} of unimodular Bianchi algebras B${\mathbf 2}$,
B${\mathbf 6_0}$, B${\mathbf 7_0}$, B${\mathbf 8}$, B${\mathbf 9}$ (see Table 1).
For all of these algebras we have calculated explicit forms of
generalized frame fields and checked that they satisfy relations
\eqref{Lieder EAEB}.

Beside that we have obtained five inequivalent generalized algebras
\eqref{eq:Enalgebra2} corresponding to the Bianchi algebras
B${\mathbf 3}$, B${\mathbf 4}$, B${\mathbf 5}$, B${\mathbf 6_a}$,
B${\mathbf 7_a}$. Their products are given by relations
\eqref{BianchiEn2}. In this case generalized frame fields
given by \eqref{EAI} must be twisted by matrix \eqref{twistmtx} to
satisfy relations \eqref{Lieder EAEB twist} for the generalized algebra.

As we have complete classification of algebras ${\mathcal E}_3$ they represent analogs  of Manin triples 
obtained for $n=3$ in \cite{hlasno:MT}. 
In case we want to study non-Abelian U-dualities
or pluralities we have to look for their mutual relation given by
linear \tfn s, more precisely, representations of the group
$E_3=SL(3)\times SL(2)$. Unfortunately, it seems that no such
relations exist. Certainly there is no linear \tfn\  between
algebras \eqref{abelianEn} and \eqref{BianchiEn} or
\eqref{BianchiEn2} as the former ones are Lie and the latter not.
Attempts to find a linear \tfn s inside these two classes by brute
force failed as well. It seems that the only possibility to find
non-Abelian U-dualities or pluralities for six-dimensional Leibniz
algebras is a further generalization of the form
\eqref{eq:Enalgebra2}.

\end{document}